\documentclass[11pt,aslarticle,xperspectives,bibay3]{asl}
\usepackage{listings}
\usepackage{bk0,bk1,bm}
\usepackage[mathscr]{euscript}
\usepackage[matrix,arrow]{xy}
\xyoption{curve}
\CompileMatrices
\usepackage[english]{babel}
\usepackage{geometry}   
\usepackage{epic}
\usepackage{graphicx}
\usepackage{proof}
\usepackage{color}
\usepackage{color}
\usepackage{robustmultind}

\newcommand\redalpha{\mathrel{\rightarrow_{\alpha}}}
\newcommand\redmu{\mathrel{\rightarrow_{\mu}}}
\newcommand\redmucl{\mathrel{\rightarrow_{[\mu]}}}
\newcommand\redmualpha{\mathrel{\rightarrow_{\mu/\alpha}}}
\newcommand\redmualphastar{\mathrel{\rightarrow^{\ast}_{\mu/\alpha}}}
\newcommand\redmua{\mathrel{\rightarrow_{\mu'}}}
\newcommand\redmustar{\mathrel{\rightarrow^{\ast}_{\mu}}}
\newcommand\redmuastar{\mathrel{\rightarrow^{\ast}_{\mu'}}}
\newcommand\tamua{\type_{\mu'}}
\newcommand\tamucl{\type_{[\mu]}}
\newcommand\tmua{\type_{\mu'}}
\newcommand\prmua{\vdash_{\mu'}}
\newcommand\prmuacl{\vdash_{[\mu']}}
\newcommand\weakmualphaeq{=_{\mu/\alpha}}
\newcommand\weakeqcl{=_{[\mu]}}
\newcommand\redst{\mathrel{\rightarrow^{\ast}_{st}}}
\newcommand\redstl{\mathrel{\leftarrow^{\ast}_{st}}}
\newcommand\alphaarrow{\vec{\alpha}}
\newcommand\alphaiarrow{\vec{\alpha_i}}
\newcommand\alphaeenarrow{\vec{\alpha_1}}
\newcommand\alphatweearrow{\vec{\alpha_2}}
\newcommand\betaiarrow{\vec{\beta_i}}
\newcommand\betaeenarrow{\vec{\beta_1}}
\newcommand\betatweearrow{\vec{\beta_2}}
\newcommand\betaarrow{\vec{\beta}}
\newcommand\restricted{\!\!\upharpoonright\!\!}
\newcommand\eqalpha{\mathrel{=_{\alpha}}}
\newcommand\SC{{\mbox {SC}}}

\begin{document}
\newcount\mscount
\renewcommand\labelx[1]{\label{#1}}
\renewcommand\refx[2]{\ref{#1}}
\renewcommand\partt[1]{\part{#1}}
\renewcommand\ldots\cdots
\renewcommand\hb[1]{{#1}}
\renewcommand\Red[1]{{#1}}
\renewcommand\bchh{{}}
\renewcommand\echh{{}}
\renewcommand\hbweg[1]{{}}
\pagenumbering{arabic}

{
%

\renewcommand{\TT}{{\cal A}}
\renewcommand{\SS}{{\cal B}}

\newcommand{\testo}[1]{~\mbox{#1}~}
\newcommand{\profinites}{\mathcal{CU}}    
\newcommand{\TypesOver}[1]{\mathcal{S}(#1)}
\newcommand{\TATypes}{{\sf T} \!\!\! {\sf T}}
\newcommand{\assstrongapp}{\vdash_{\lambda \mu^{\star\infty}}}  
\newcommand{\strongapp}{{\lambda \mu^{\star\infty}}}

\newcommand{\ato}{\to}
\newcommand{\primeset}[1]{||#1||}
\newcommand{\notrel}{\not\!\!\rel}
\newcommand{\subty}{{\cal S}}
\newcommand\solstar[1]{\cour{sol}^{\star}({#1})}
\def\T{\type}
\renewcommand{\to}{\! \rightarrow \!}
\def\Tcons{{\sf C} \! \! {\sf C}}     

\def\types{\type}
\newcommand{\Types}{{\type}^{\!\mu}}
\newcommand\typems{\type^{\star}_\mu}
\newcommand{\Typesplus}{{\type}_{\mu}^+}
\newcommand{\SimpleTypes}{\type}
\newcommand{\ATypes}[1]{{\type}_{\small #1}}
\newcommand{\CTypes}[1]{{\type}_{c_1,\ldots,c_{#1}}}
\newcommand{\contrTypes}{{\type}^{\mu c}}
\newcommand{\reg}{\rm{\hb{R}}}
\newcommand{\mt}[2]{\mu #1 . #2}
\newcommand{\mtu}[2]{#2[#1:=\mu #1.#2]}
\newcommand{\GTypes}{{\sf T \!\! \! T}}
\newcommand{\atset}{{\bf A}}
\newcommand{\undef}{\bottom}

\newcommand{\stc}{\EE}                     

\newcommand{\srcr}{{\cal R}^\diamond}

\newcommand{\inv}[1]{{#1}^{\rm inv}} 
\newcommand{\simple}[1]{{\inv{#1}}^-}

\newcommand{\sre}[1]{(#1)^{\star}}
\newcommand{\princ}[1]{\overline{#1}}

\newcommand{\Pstc}[1]{\stc_{#1}}
\newcommand{\PG}[1]{\Gamma_{#1}}
\newcommand{\PT}[1]{a_{#1}}


\newcommand{\Kcomb}{\mbox{\bf K}}
\def\two{\mbox{\bf two}}
\newcommand{\auto}{\lambda x . x  x}
\newcommand{\Tauto}[2]{\mt{#1}{(#1 \to #2)}}
\newcommand{\fauto}[1]{\lambda x . #1 (x  x)}
\newcommand{\fixpoint}{\lambda f.(\fauto{f})(\fauto{f})}
\newcommand{\Tdelta}{T_{\Delta}}
\newcommand{\typeof}[1]{\tau(#1)}
\newcommand{\fail}{\mbox{FAIL}}
\renewcommand{\int}{\mbox{\bf int}}



\newcommand{\prfmuw}{\vdash_{\mu}}
\newcommand{\prfmus}{\vdash_{\mu^{\star}}}
\newcommand{\prfCw}[1]{\vdash_{#1}}
\newcommand{\prfCs}[1]{\vdash_{#1}^{\star}}
\newcommand{\prfCsAK}{\vdash_{\mu^{\star} {\AK}}}
\newcommand{\prfCwinv}[1]{\vdash_{#1}^{\rm inv}}

\newcommand{\eqmu}{\vdash_{\mu}}   
\newcommand{\eqC}[1]{\vdash_{#1}}
\newcommand{\eqmuw}{\vdash_{\mu\sim}}
\newcommand{\eqCw}[1]{\vdash_{#1}}
\newcommand{\eqCwinv}{\vdash^{\star}_{\sim}}
\newcommand{\eqmus}{\vdash_{{\rm BH}}}
\newcommand{\eqmusAK}{\vdash_{\mu^{\star} {\AC}}}
\newcommand{\eqCs}{\vdash_{\approx}}

\newcommand{\weqsys}[1]{\vdash_{#1}}
\newcommand{\seqsys}[1]{\vdash_{{#1}^{\star}}}


\newcommand{\assR}{\vdash_{\lambda\mu^{\sharp}}}
\newcommand{\assmu}{\vdash_{\lambda\mu^-}}
\newcommand{\asseq}{\assweak}
\newcommand{\assweak}{\vdash_{\lambda\mu}}
\newcommand{\assweakplus}{\vdash_{\lambda\mu +}}
\newcommand{\assstrong}{\vdash_{\lambda\mu^{\star}}}

\newcommand{\assTA}[1]{\vdash_{#1}}

\newcommand{\assCR}[1]{\vdash_{\lambda {#1}^{\sharp}}}
\newcommand{\assCeq}[1]{\assCweak{#1}}
\newcommand{\assCweak}[1]{\vdash_{\lambda #1}}
\newcommand{\assCstrong}[1]{\vdash_{\lambda {#1}^{\sharp}}}
\newcommand{\assCgen}{\vdash_{\lambda \geneq}}

\newcommand{\assChTA}[1]{\vdash_{\lambda {#1}^{\rm Ch}}}
\newcommand{\assCh}{\vdash_{\lambda_\mu^{\rm Ch}}}
\newcommand{\assChweak}{\vdash_{\lambda_\mu{\rm Ch}}}
\newcommand{\assChweako}{\vdash_{\lambda\mu{\rm Ch}_0}}
\newcommand{\assChstrong}{\vdash_{\lambda\mu^{\star} {\rm Ch}}}
\newcommand{\assChR}{\vdash_{\lambda\relmu{\rm Ch}}}
\newcommand{\assCCh}[1]{\vdash_{\lambda{\rm Ch}{\Ceq{#1}}}}
\newcommand{\assCChweak}[1]{\vdash_{\lambda{\rm Ch}{\weakCeq{#1}}}}
\newcommand{\assCChstrong}[1]{\vdash_{\lambda{\rm Ch}{\strongCeq{#1}}}}

\newcommand{\assdBTA}[1]{\vdash_{\lambda {#1} \text{-dB}}}
\newcommand{\assdB}{\vdash_{\lambda\mu{\rm dB}}}
\newcommand{\assdBweak}{\vdash_{\lambda\mu{\rm dB}}}
\newcommand{\assdBweako}{\vdash_{\lambda\mu{\rm dB}_0}}
\newcommand{\assdBstrong}{\vdash_{\lambda\mu^{\star} {\rm dB}}}
\newcommand{\assdBR}{\vdash_{\lambda\relmu{\rm dB}}}
\newcommand{\assCdB}[1]{\vdash_{\lambda{\rmdB}{\Ceq{#1}}}}
\newcommand{\assCdBweak}[1]{\vdash_{\lambda{\rm dB}{\weakCeq{#1}}}}
\newcommand{\assCdBstrong}[1]{\vdash_{\lambda{\rm dB}{\strongCeq{#1}}}}


\newcommand{\Equiv}{\mbox{(equal)}}
\newcommand{\ruleCweak}{\mbox{($\sim_{\cal C}$)}}
\newcommand{\ruleCstrong}{\mbox{($\approx_{\cal C}$)}}

\newcommand{\rulefold}{\mbox{(fold)}}
\newcommand{\ruleunfold}{\mbox{(unfold)}}
\newcommand{\ruleweak}{\mobx{($\weakeq$)}}
\newcommand{\rulestrong}{\mobx{($\strongeq$)}}
\def\ax{\mbox{(axiom)}}
\newcommand{\const}{\mbox{(const)}}
\newcommand{\eqaxmu}{\mbox{($\mu$-eq)}}
\newcommand{\eqaxsr}{\mbox{($\sr$-eq)}}
\newcommand{\eqaxC}{\mbox{(eq)}}
\newcommand{\ident}{\mbox{(ident)}}
\newcommand{\reflex}{\mbox{(symm)}}
\newcommand{\trans}{\mbox{(trans)}}
\newcommand{\rarrow}{\mbox{($\into$-cong)}}
\newcommand{\rmu}{\mbox{($\mu$-cong)}}
\newcommand{\drmu}{\mbox{($\dm$-cong)}}
\newcommand{\intro}{\mbox{($\into$I)}}
\newcommand{\muintro}{\mbox{($\mu$I)}}
\newcommand{\invert}{\mbox{(inv)}}
\newcommand{\elim}{\mbox{($\into$E)}}
\newcommand{\fullrule}[4]{{\textstyle (#1)}~{\frac{\textstyle #2}
                          {\textstyle #3}}~\mboxrm{#4}}
\newcommand{\myrule}[3]{{\textstyle (#1)}~{\frac{\textstyle #2}
                          {\textstyle #3}}}
\newcommand{\srule}[2]{\bpt #1\hence #2 \ept}
\newcommand{\rulemuuno}{(R_{\EE}^{\rm uf})}
\newcommand{\rulemudue}{(R_{\EE}^{\rm fu})}


\def\lm{\l\mu}
\newcommand{\lambdamuR}{\lambda  \relmu}
\newcommand{\lambdamuweak}{\lambda \mu}
\newcommand{\lambdamuweakplus}{\lambda \mu^+}
\newcommand{\lambdamustrong}{\lambda \mu^{\star}}

\newcommand{\lambdaCR}[1]{(\lambda {#1}^{\sharp})}
\newcommand{\lambdaCeq}{\lambdaCweak{?}}
\newcommand{\lambdaCweak}[1]{\lambda #1}
\newcommand{\lambdaCstrong}[1]{\lambda {#1}^{\star}}
\newcommand{\lambdaCgen}{\lambda \geneq}

\newcommand{\lambdaTA}[1]{(\lambda #1)}

\newcommand{\ChurchTA}[1]{\lambda^{#1\mbox{-}\ch}}
\newcommand{\lmuChurch}{\lambda_\mu^{\ch_0}}
\newcommand{\lmuChweak}{\lambda_\mu^{\ch}}
\newcommand{\lAChurch}{\lambda^{\AA\mbox{-}\ch_0}}
\newcommand{\lAChweak}{\lambda^{\AA\mbox{-}\ch}}
\newcommand{\lmuChstrong}{\lambda_{\mu^{\star}}^{\ch}}
\newcommand{\lmuChurchR}{\lambda_\relmu^\ch}

\newcommand{\lCChurch}[1]{(\lambda^\ch\Ceq{#1})}
\newcommand{\lCChurchweak}[1]{(\lambda^\ch\weakCeq{#1})}
\newcommand{\lCChurchstrong}[1]{(\lambda^\ch\strongCeq{#1})}

\newcommand{\dBTA}[1]{\bl_=^{#1,\db}}
\newcommand{\lmudB}{(\lambda\mu \mbox{-} \mbox{dB}_0)}
\newcommand{\lmudBweak}{(\lambda\mu \mbox{-}  \mbox{dB})}
\newcommand{\lmudBstrong}{(\lambda\mu^{\star}  \mbox{-} \mbox{dB})}
\newcommand{\lmudBurchR}{(\lambda\relmu \mbox{-} \mbox{dB})}

\newcommand{\lCdB}[1]{(\lambda\mbox{dB}\Ceq{#1})}
\newcommand{\lCdBweak}[1]{(\lambda\mbox{dB}\weakCeq{#1})}
\newcommand{\lCdBstrong}[1]{(\lambda\mbox{dB}\strongCeq{#1})}


\newcommand{\eqmuwsys}{(\mu)}
\newcommand{\eqmussys}{{\rm (BH)}}
\newcommand{\eqCwsys}[1]{(#1)}
\newcommand{\eqCssys}[1]{({#1}^{\star})}

\newcommand{\eqmusAKsys}{(\mu ^{\star}_{\rm AC})}

\newcommand{\eqCwinvsys}[1]{({#1}^{\rm inv})}
\newcommand{\inveq}[1]{=_{#1}^{\rm inv}}


\newcommand{\weakeq}{\sim}
\newcommand{\strongeq}{{=_{\mu}^{\star}}}
\newcommand{\weakCeqinv}[1]{=_{#1}^{\rm inv}}
\newcommand{\strongCeq}[1]{=_{#1}^{{\star}}}
\newcommand{\geneq}{\simeq}
\newcommand\tr[1]{#1^{{\star}}}

\newcommand{\relmu}{\mu^{\sharp}}
\newcommand{\relC}[1]{{#1}^{\sharp}}

\newcommand{\wec}{=}
\newcommand{\vwec}{=}

\newcommand{\Cwequiv}{\sim}


\newcommand{\TAold}[1]{\corners{#1}}
\newcommand{\TA}[1]{\corners{#1,\to}}
\newcommand{\STAE}[1]{\type_{#1}}
\newcommand{\TAE}[1]{\TT_{#1}}

\newcommand{\Qset}[2]{{#1}/_{\! {#2}}}
\newcommand{\Eqclass}[2]{[#1]_{#2}}


\newcommand{\imply}{\Longrightarrow}
\newcommand{\subgen}[4]{[#1_{#3} := #2_{#3}, \ldots, #1_{#4} := #2_{#4}]}
\newcommand{\redbeta}{\longrightarrow_{\beta}}
\newcommand{\redeta}{\longrightarrow_{\eta}}
\newcommand{\redbetaeta}{\longrightarrow_{\beta\eta}}
\newcommand{\ded}{{\sf D}}

\newcommand{\union}{\cup}
\newcommand{\ih}{induction hypothesis}
\newcommand{\bottom}{\perp}
\renewcommand{\sp}{~}
\def\sep{\sp | \sp}
\newcommand{\SAT}{{\tt SAT}}
\newcommand{\semto}{\Rightarrow}

\newcommand{\ec}{\! = \!}

\newcommand{\calY}{{\cal Y}}
\newcommand{\calc}{{\cal C}}
\newcommand{\calI}{{\cal H}}
\newcommand{\calE}{{\cal E}}
\newcommand{\calA}{{\cal A}}
\newcommand{\calK}{{\cal K}}

\newcommand{\bfA}{{\bf A}}
\newcommand{\bfI}{{\bf I}}

\newcommand{\simrec}{simultaneous recursion}
\newcommand{\acapo}{\newline \noindent}


\newcommand{\assume}{\mbox{(hyp)}}
\newcommand{\coind}{\mbox{(coind)}}
\newcommand{\noncontract}{\mbox{(triv)}}
\newcommand{\AK}{\mbox{({\rm AC*})}}
\def\AC{\mbox{(AC)}}
\newcommand{\deruno}{\mbox{(der1)}}
\newcommand{\derdue}{\mbox{(der2)}}
\def\k{{\kappa}}
\newcommand{\semeqC}[2]{=^{#1}_{#2}}
\newcommand{\semeq}[1]{=_{#1}}
\newcommand{\clos}{{\cal SC}}
\def\redstar{\to^{\star}}
\newcommand{\trs}[1]{{\rm TRS}^{-1}(#1)}
\newcommand{\trscr}[1]{\mbox{TRS}^\diamond(#1)}
\newcommand{\trsplus}{\trscr}            
\newcommand\trsi{\mbox{Trs}^{-1}}
\newcommand{\ov}{\overline}


\newcommand{\eqCsys}{(=)}
\newcommand{\eqmusys}{??}
\newcommand{\lambdamueq}{\lambdamuweak}
\newcommand{\mueq}{\weakeq}


\newcommand{\valof}[2]{{\cal I}\sem{#1}_{#2}}
\newcommand{\paramvalof}[3]{{\cal I}^{#3}\sem{#1} #2}
\newcommand{\valoftree}[2]{{\cal T}\sem{#1}_{#2}}
\newcommand{\valofterm}[2]{\sem{#1}_{#2}}
\newcommand{\valoftype}[2]{{\sem{#1}_{#2}}}
\newcommand{\emb}{\lhd}
\renewcommand{\into}{\rightarrow}
\renewcommand{\dom}{{\rm dom}}
\newcommand{\eqdef}{\eqdf}
\newcommand{\supremum}{{\bf sup}}
\renewcommand{\lmu}{\lambda\mu}
\newcommand{\image}[1]{{\rm im}(#1)}
\newcommand{\universe}{{\bf V}}
\newcommand{\fix}{{\bf fix}}
\renewcommand{\bm}[1]{\mbox{\boldmath $#1$}}
\newcommand{\fold}[1]{{\it fold}_{#1}}
\newcommand{\unfold}[1]{{\it unfold}_{#1}}
\newcommand{\lub}{\bigsqcup}
\newcommand{\sierpinski}{{\sf O}}
\newcommand{\appvalof}[3]{{\cal I}^{#1}\sem{#2}_{#3}}
\newcommand{\rappvalof}[3]{{\cal T}^{#1}\sem{#2}_{#3}}
\newcommand{\nskip}[1]{}
\newcommand{\Pomega}{{\cal P}\omega}
\newcommand{\finsets}[1]{{\cal P}_{\scriptstyle\rm fin}(#1)}
\newcommand{\typecarrier}[1]{\mathcal{M}(#1)}
\newcommand{\grph}{{\bf graph}}
\newcommand{\unk}[1]{{\rm Dom}(#1)}     
\newcommand{\cald}{{\cal D}}
\newcommand{\cale}{{\cal E}}
\newcommand{\height}[1]{\| #1 \|}
\newcommand{\length}[1]{\left| #1 \right|}
\newcommand{\truncation}[2]{{#1}_{| #2}}
\newcommand{\rattreeOmega}{{\bf Tr}^R_{\Omega}}
\newcommand{\inftreeOmega}{{\bf Tr}^{\rm inf}_{\Omega}}
\newcommand{\metalambda}{\lambda \! \! \! \lambda}
\newcommand{\closures}{\mathcal{V}}

\newcommand\Tover[1]{\type({#1})}
\newcommand\typeplus{\type_{1,+,\times}}
\newcommand\typea{\Tover{\Tatom}}
\newcommand\ise{=_{\EE}}
\newcommand\isr{=_{\sr}}
\newcommand\XX{{\mathcal X}}
\newcommand\tterm[2]{\cour{texp}_{#1}[#2]}
\newcommand\defdby{\:\:\!=}
\newcommand\wordt{\:\!=}
\newcommand\TTT{\widetilde{\TT}}
\newcommand\Diag[1]{\cour{Diag}_{#1}}
\newcommand\hn{h^{\natural}}
\newcommand\LTT{{\cal L_{\TT}}}
\newcommand\rat{\comb{Q}}
\newcommand\aap[1]{#1}

\def\change{\renewcommand\newblock{\\}}
\newcommand{\inftree}{\inftrees}
\newcommand\neutral{\natural}


\newcommand\TQE[2]{#1/#2}
\newcommand{\weakCeq}[1]{=_{#1}}
\newcommand\TTA{\widetilde{\types}}


\newcommand\sol[1]{\cour{sol}({#1})}
\newcommand\sols[1]{\cour{sol}^{\star}({#1})}
\newcommand\hnat{h^{\natural}}

\newcommand\TTC{\TT_{\CC}}
\newcommand\SSC{\SS_{\CC}}
\newcommand{\ext}[2]{{#1[#2]}}
\newcommand{\extp}[2]{{#1[#2]}}
\newcommand{\exts}[2]{#1(#2)^{\star}}
\renewcommand{\mod}[2]{{#1}/{#2}}
\newcommand{\dd}[1]{}
\newcommand{\less}{\sqsubseteq}
\renewcommand{\strongCeq}[1]{=^{\star}_{#1}}
\newcommand{\trunc}[2]{(#1)_{#2}}
\newcommand{\trunx}[2]{\left. {#1}\right|_{#2}}
\newcommand{\vc}[2]{{#1}_1,\ldots,{#1}_{#2}}
\newcommand{\vecX}{{\vec{X}}}
\newcommand{\vecY}{{\vec{Y}}}
\newcommand\fto\to
\renewcommand\TTO{\type_o}

\renewcommand{\Types}{{\type_{\mu}}}
\renewcommand{\weakeq}{{\; =_{\mu} \,}}
\newcommand{\nweakeq}{{\; \neq_{\mu} \,}}
\newcommand{\nstrongeq}{{\neq_{\mu}^{\star}}}
\newcommand{\tamu}{\T_{\mu}}    
\newcommand{\tamustar}{\TT_{\mu}^{\star}}

\newcommand\weq\weakCeq
\newcommand\treq{\tr{=}}
\newcommand\streq[1]{{\;\treq_{#1}\;}}
\newcommand\sreq[1]{{{=}^{\star}_{#1}}}

\newcommand\appeq[1]{{\; =_{\mu}^{#1} \,}}
\newcommand\srmin{\sr^{\!^-}}
\newcommand{\tmax}[2]{{(#1)}^{{\star}}_{#2}}
\newcommand{\tmap}[2]{{(#1)}^{\star}_{#2}}
\newcommand\stmu[1]{(#1)^{\star}_\mu}



\newcommand{\old}[1]{}
\newcommand{\new}[1]{#1}
\renewcommand{\T}  {\type}    
\newcommand{\sta}[2]{#1/\!#2}
\newcommand{\idrel}{=}
\renewcommand{\TTO}{\T_\om}
\newcommand{\Tvars}{{\mathbb V}}   
\newcommand{\muT}{\Types}     
\newcommand{\mutypes}{\Types}
\renewcommand{\Atoms}{\Tatom}
\newcommand{\ccn}[1]{{\color{blue}{#1}}}   
\newcommand{\simplecenter}[1]{\bsub\item[]\begin{center} #1 \end{center}\esub}

\newcommand{\scent}[1]{\simpecenter{#1}}

\newcommand{\ata}{\struct{|\AA|,\to}}



\newif\ifmarc\marctrue

\definecolor{newc}{rgb}{0,.5,.5}
\newcommand{\mar}[1]
{\ifmarc{\color{newc}{#1}}\else{#1}\fi}
\newcommand{\marc}[1]{\ifmarc{\footnote{{\mar{Mario's comm.:#1}}} }\else\fi}
\newcommand{\mc}[1]{\begin{color}{red} {#1} \end{color}}

\newcommand\KD{\KK(\dD)}

\newcommand{\restr}{\upharpoonright}
\newcommand{\inl}{\mathtt{inl}}
\newcommand{\inr}{\mathtt{inr}}
\newcommand{\arrow}[3]{
#1 \stackrel{#2}{\longrightarrow} #3}

\newcommand\lAs{{\bl\AA}}
\newcommand\laes{{\bl\AA}}
\newcommand{\lR}{{\bl_=^{\sr}}}
\newcommand{\lRs}{{\bl{\sr}}}
\newcommand\lrel{{\bl_=^{\type/\rel}}}
\newcommand\lrels{{\bl\rel}}
\newcommand\lE{{\bl_=^{\type/\EE}}}
\newcommand\lEst{{\bl_=^{\type/{\EE\st}}}}
\newcommand\lEsts{{\bl{\EE\st}}}

\newcommand\lEs{{\bl\EE}}

\newcommand\pra{\mathbin{\pr_{\AA}}\,}
\newcommand\prel{\mathbin{\pr_{\type/\rel}}\,}
\newcommand\prE{\mathbin{\pr_{\bl\EE}}\,}
\newcommand\prsr{\mathbin{\pr_{\type[\sr]}}\,}
\newcommand\prmu{\mathbin{\pr_{\type_\mu}}\,}

\newcommand\eqrule{\mbox{(equiv)}}
\newcommand{\Lta}{{\L_=^{\AA,\ch}}}
\newcommand\laech{{\bl_=^{\AA,\ch}}}
\newcommand\laecho{{\bl_=^{\AA,\ch_0}}}
\newcommand\LAch{{\L_=^{\AA,\ch}}}
\newcommand\LAcho{{\L_=^{\AA,\ch_0}}}
\newcommand\we{\weakeq}
\newcommand\ismu{{=_\mu^*}}
\newcommand{\tmu}[1]{\ttmap{#1}{\mu}}

\def\ln{\ar@{-}} \def\ld{\ar@{.}}

\newcommand\tamum{\tamu/\!=_\mu}
\newcommand\typamu{\typam/\!=_\mu}
 \title {Weak $\mu$-equality is decidable}
\author{Wil Dekkers, Radboud University, Nijmegen, The Netherlands}
\maketitle
\newenvironment{Abstract} {\begin{center}\textbf{Abstract}\end{center} \begin{quote}} {\end{quote}}
\begin{Abstract}
In this paper we consider the set $\tamu$ of $\mu$-types, an extension of the set $\T$ of simple types freely generated from a set $\Atoms$ of atomic types and the type constructor $\to$ , by a new operator $\mu$, to explicitly denote solutions of recursive equations like $A\weakeq A\to \beta.$ We show that this so-called weak $\mu$-equality for $\mu$-types is decidable by defining a derivation system for weak $\mu$-equality based on standard reduction for $\mu$-types such that the number of nodes in a derivation tree for $\ixsrt{A \weakeq B}$ is bounded as a function of $A,B.$ We give two proofs. One for decidability of $\weakeq$ for $\alpha$-equivalence classes of $\mu$-types and one for decidability of $\weakmualphaeq$ for $\mu$-types themselves. Both proofs are straightfroward and elementary.
\end{Abstract}
\newenvironment{Introduction} {\begin{center}\textbf{Introduction}\end{center} \begin{quote}} {\end{quote}}
\begin{Introduction}
In Cardone and Coppo[1] a proof method is given to show decidability of weak $\mu$-equality, using standard reduction and a special purpose proof system. It turned out recently that their treatment of $\alpha$-conversion was not completely correct. Nethertheless their result is correct and  in Endrullis, Grabmayer, Klop and van Oostrom[2] three proofs for the decidability are given. Two of the given proofs are inspired by the Cardone-Coppo proof strategy. The third proof uses the theory of regular languages and is totally different. This paper is based on the first two proofs of Endrullis et al. These proofs are ingenuous but they are also technical and intricate. Our aim is to give simpler and more straightforward proofs.
\newline In their first proof Endrullis et al. work with $\mu$-types themselves, not with $\alpha$-equivalence  classes of $\mu$-types. The proof constructs a decision procedure for equality of types based on standard reduction for types. In that proof they need the theory of $\alpha$-avoiding developed by van Oostrom[3]. In the second proof they work with $\alpha$-equivalence classes  and the corresponding reduction relation $\redmu$ for classes and they show the decidability of $\weakeq$. Our proof is more or less the other way round. Based on the first proof of Endrullis et al. of decidabiliy of equality for types we give a proof of decidability of $\weakeq$ for classes of types and then from that proof we derive a first order proof of decidability of equality for types where the theory of $\alpha$-avoiding is not needed. Both proofs are straightforward and elementary.
\newline So in our first proof $\tamu$ is he set of $\alpha$-equivalence classes of types and  $A, B$ stand for classes of types. We start by defining a set $\tamua$, containing $\tamu$, of so-called annotated ( classes of ) $\mu$-types and a reduction $\redmua$ : $\tamua \to \tamua$ that is closely related with standard reduction on $\tamu$. Then we define a derivation system $\prmua$ for equality of elements of $\tamua$ such that for $A,B\in \tamu$ one has 
$$\prmua A = B \Leftrightarrow A \mbox{ and } B \mbox{ have a common standard reduct.}$$
We show for $a\in\tamua$ that $\SC(a) \equiv \{b\in\tamua | a\redmuastar b\}$ is bounded as a function in $a$ and also that for $ a = b$ occurring in a derivation tree for $\prmua A = B$ we have $a\in\SC(A)$ and $b\in \SC(B).$ So  the number of nodes occurring in a derivation tree for  $\prmua A = B$ is bounded as a function in $A,B.$ So $\prmua A = B$ is decidable and hence $A\weakeq B$ is decidable, because of  the fact that 
$$A\weakeq B  \Leftrightarrow A \mbox{ and } B \mbox{ have a common standard reduct.}$$

Then in part 2 from this result we derive easily that $\weakmualphaeq$ on  types themselves can be decided by first order means via a decision procedure for $\weakmualphaeq$ on types. 

\end{Introduction}
\section{ A decision procedure for weak $\mu$-equality via $\alpha$-equivalence classes of types}
\bdf[Of $\tamu$ and $\redmu$]\label{tamu and redmu}
\bsub\firstitem  Let $\Tatom=\Tatom_\infty$ be an infinite set of type
atoms $\alpha,\beta,\gamma\ldots$  considered as type variables for the purpose of binding and
substitution.The set of $\mu$-types, notation $\tamu$, is defined as follows.
$\;\;\;\tamu = \alpha\mid A_1\to A_2\mid\mu\alpha.A_1.$
\item On $\tamu$ we define the notion of
$\alpha$-reduction and $\alpha$-conversion via the contraction rule
$$\mu\alpha.A\mapsto_\a \mu\a'.A[\a:=\a'], \text{ provided
}\a'\notin\FV(A).$$ The relation $\redalpha$ is the least
compatible relation containing $\mapsto_\a$ and $\eqalpha$ is  the least congruence containing
$\mapsto_\a$. 
\item Define on $\tamu$ a notion of $\mu$-reduction via the contraction rule $\mapsto_{\mu}$
$$\mu\beta.A_1\mapsto_{\mu} A_1[\beta:=\mu\beta.A_1]$$ The relation $\redmu \subseteq \tamu \times \tamu$ is the compatible closure of $\mapsto_{\mu}.$ That is
\beqn
&&A_1\redmu B_1\imp\mu\beta.A_1\redmu\mu\beta.B_1;\\
&&A_1\redmu A'_1\imp (A_1\to A_2)\redmu (A'_1\to A_2);\\
&&A_2\redmu A'_2\imp (A_1\to A_2)\redmu (A_1\to A'_2).
\eeqn
\item $\redmustar$ denotes zero or more $\redmu$ steps and $\weakeq$ is the least congruence containing $\mapsto_{\mu}.$
\item In this first part we work with $\alpha$-equivalence classes of types and we asume that all bound and free variables are different. Especially in $\mu\alpha_1 ...\mu\alpha_n.A$ all $\alpha_i$ are different. 
 \esub \edf

\bdf Let $[A]$ denote the $\alpha$-equivalence class of A and let $\tamucl$ denote the set of $\alpha$-equivalence classes. $\redmu$ and $\weakeq$ can easily be lifted to relations $\redmucl$ and $\weakeqcl$ on $\tamucl.$ In part 2 of this paper we will consider $\tamu,\;\redmu$ and $\weakeq$ but in this part 1 we work with $\tamucl,\;\redmucl$ and $\weakeqcl$ and we simply write  $\tamu,\;\redmu$ and $\weakeq.$ Also we write $A$ for $[A].$ 
\edf

\bdf\label{standardreduction}[Of standard reduction $\redst$]\\
 (a) $A \redst A$\\
(b) If $A_1[\beta :=\mu\beta.A_1]\redst B,$ then the composition $\mu\beta.A_1\to A_1[\beta :=\mu\beta.A_1]\redst B$ is a standard reduction.\\
(c) If $A_1\redst B_1,$ then the induced reduction $\mu\beta.A_1\redst\mu\beta. B_1$ is a standard reduction.\\
(d) If $A_1\redst A'_1$ and  $A_2\redst A'_2,$ then each reduction $(A_1\to A_2)\redmustar (A'_1\to A'_2)$ generated by interleaving is a standard reduction.
\edf

\begin{example}\label{streductions}
\bsub\fit Let $A \equiv\mu\alpha\mu\beta.\alpha.$ Then $A\redmu\mu\beta_1.A\redmu\mu\beta_1\mu\beta_2.A\redmu\ldots$ is a standard reduction.
\item Let $B\equiv\mu\alpha.\alpha\to\alpha.$ Then $B\redmu(B\to B)\redmu(B\to B)\to B\redmu\ldots$ is a standard reduction.
\item Let $C\equiv\mu\alpha\mu\beta\mu\gamma.\alpha.$ Then $C\redmu\mu\alpha\mu\gamma.\alpha\redmu\mu\gamma_1(\mu\alpha\mu\gamma.\alpha)$ is not a standard reduction. 
\esub
\end{example} 

\brem\label{facts} We have the well known facts.
\bsub \item The relation $\redmu$ is Church Rosser (CR);
\item If $A \redmustar B$ then there is a standard reduction $A\redst B.$
\esub
\erem

We illustrate (ii) in the above remark by an example.
\begin{example}\label{examplestreduction}
Let $A=\mu\alpha\mu\beta.C[\alpha,\beta].$ Here $C[.,.]$ is a so-called context. The variables $\alpha$ and $\beta$ can occur several times in it. Now let $B=\mu\alpha.C[\alpha,\mu\beta.C].$ Then
$$A\redmu B=\mu\alpha.C[\alpha,\mu\beta.C[\alpha,\beta]]\redmu C[B,\mu\beta.C[B,\beta]].$$
We can make this two-step reduction into a standard reduction as follows.
$$A\redmu\mu\beta.C[A,\beta]\redmu C[A,\mu\beta.C[A,\beta]]$$
and now reduce each occurrence of $A$ to $B$, from left to right.
\end{example}

Let for the moment $A\downarrow B$ denote that $A$ and $B$ have a common standard reduct. Our aim is to show that $A\weakeq B$ is decidable. By (ii) of the above remark we are ready if we show that $A\downarrow B$ is decidable. However a direct proof of that is not so easy because in general a type $A$ may have infinitely many standard reducts as is shown by (i) and (ii) of Example \ref{streductions}.

We remedy the two problems given by the example by changing the notion of standard reduction (and also the set of types)  roughly as follows. In clause (c) of the notion of standard reduction we add the condition $\beta\in FV(A_1).$ Then the number of $\mu\beta_i$ as in (i) of the  example  is bounded . Moreover clause (d) is changed into $(A_1\to A_2)\to A_i\;\;\;\;i = 1,2$ to remedy the problem shown in (ii).

This is formalised as follows.

\bdf[Of the set of annotated types $\tmua$]\label{annotatedtypestmua}
\bsub\fit The set of annotated types $\tmua$ is defined by
$\tmua \equiv \{(\mu\alpha_1).....(\mu\alpha_n)A\mid A\in\tamu,\alpha_i\in FV(A),\alpha_i\not=\alpha_j\mbox{ if } i\not= j\},$
\item We consider the annotated types modulo $\alpha$-equivalence as follows \\
$(\mu\alpha_1).....(\mu\alpha_n)A\equiv(\mu\alpha'_1).....(\mu\alpha'_n)A[\alpha_i :=\alpha'_i]$  $(\alpha'_i\notin \FV(A)\cup\BV(A)).$ And implicitly $\alpha'_i\not=\alpha'_j\mbox{ if } i\not= j\},$  
\esub\edf

\brem\label{overtmua}
\bsub\fit In Endrullis et al.[2] a set AnnTer$(\mu)$ is defined by AnnTer$(\mu) = \{(\mu\alpha_1).....(\mu\alpha_n)A\mid A\in\tamu\}.$ So they don't have the conditions $\alpha_i\in FV(A)$ and $\alpha_i\not=\alpha_j$ if $ i\not= j.$
Moreover they don't work modulo $\alpha$-conversion.
\item The intuition for $(\mu\alpha_1).....(\mu\alpha_n)A$ is that it stands for $\mu\alpha_1.....\mu\alpha_n.A\in \tamu$ where the $\mu\alpha_i$ at the root are frozen. This will become clear in Definition \ref{redmua}.
\item Note that $\tamu\subset\tamua.$ 
\esub
\erem

\bnnot\labelx{shorthand}
\bsub\fit We denote $(\mu\alpha_1).....(\mu\alpha_n)A$ simply as $(\mu\alpha_1...\alpha_n)A$ and mostly we write $(\mu\alphaarrow)A.$
\item $a,b,c....$denote elements of $\tamua$ and as before $A,B,C....$ stand for elements of $\tamu.$
\item If $\alphaarrow$ is a sequence of (different) type variables then $\alphaarrow\restricted  a$ denotes the sequence that arises from $\alphaarrow$ by omitting the variables that do not occur freely in $a.$  Similarly for $\alphaarrow \restricted a\cap  b.$\\
$(\mu\alphaarrow)\Box A$ stands for $(\mu(\alphaarrow \restricted A))A.$\\
If $S\subseteq \tamua$ then $(\mu\alphaarrow)\Box S = \{(\mu\alphaarrow)\Box a \mid a\in S\}.$
\item If $S\subset\tamua$ then $ |S|$ stands for the number of elements of $S.$
\item $\{\alphaarrow\}$ stands for the set of variables occurring in $\alphaarrow.$ When we write $\{\alpha_1,...,\alpha_n\}$ we always assume  $\alpha_i\not= \alpha_j$ for $i\not= j.$
\esub
\ennot

\bdf\label{redmua}(0f $\redmua : \tamua\to\tamua$ and $\SC(a)$)\bsub\item The reduction relation $\redmua : \tamua\to\tamua$ is defined as follows.
\beqn
(\mu\alphaarrow)\mu\beta.A_1&\redmua& (\mu\alphaarrow\beta)A_1\mbox{    if }\beta\in FV((A_1);\\
(\mu\alphaarrow)\mu\beta.A_1&\redmua& (\mu\alphaarrow)A_1[\beta :=\mu\beta.A_1];\\
(\mu\alphaarrow)(A_1\to A_2)&\redmua&(\mu\alphaarrow)\Box A_i\;\;\;\; i=1,2;
\eeqn
\item $\SC(a)= \{b\in\tamua\mid a\redmuastar b\}.$
\esub\edf

\brem\label{notcompatible}
\bsub\fit Note that $\redmua$ is not compatible with $\mu$ or $\to$. So for example we do not have $\mu\alpha\mu\beta.A_1\redmua\mu\alpha.A_1[\beta :=\mu\beta.A_1].$
\item Now the outline of the decidability proof is as follows.\\
--We define a derivation system $\prmua a=b$ for types $a,b\in\tamua$, closely connected with the reduction relation $\redmua$ and we show for $A,B\in\tamu$:
$$\prmua A=B\Leftrightarrow A\mbox{ and } B\mbox{ have a common standard reduct.}$$
Hence we have  $A\weakeq B\Leftrightarrow\;\prmua A=B.$\\
-- We note that for all nodes $c=d$ occurring in a derivation tree for $\prmua a=b$ we have $c\in \SC(a)$ and $d\in \SC(b.$\\
-- We show that $|\SC(a)|$ is bounded as a function in $a$. Hence the number of different nodes in a derivation tree for $\prmua a=b$ is bounded as a function in $a,b.$\\
-- We conclude that $\prmua a=b$ is decidable and hence so is $A\weakeq B$.
\esub\erem

We start by defining the derivation system $\prmua a=b$ for $a,b\in\tamua.$ We give the definition via representatives, types instead of $\alpha$-equivalence classes of types, and we show that it is invariant under $\alpha$-conversion. Hence this defines in fact  a relation on the set of $\alpha$-equivalence classes.

\bdf\label{prmua} (Of $\prmua a=b$ for $a,b\in\tamua$)
$$\bar[b]{|ll|}
\hline\hoog{1.5em}{} 
\mbox{{(axiom)}}& {\textstyle {a=b} }
                                ~~~~~\text{ if }a\eqalpha b\hoog{1.2em}\\[1em]
{\mbox{(left }}\mu{\mbox{-step)}}& \Ruled {(\mu\alphaarrow)A_1[\gamma :=\mu\gamma.A_1] = (\mu\betaarrow)B}{(\mu\alphaarrow)\mu\gamma.A_1 = (\mu\betaarrow)B} 
\\[2em]
  {\mbox{(right }}\mu{\mbox{-step)}}& \Ruled{(\mu\alphaarrow)A = (\mu\betaarrow)B_1[\delta :=\mu\delta.B_1]}{(\mu\alphaarrow)A = (\mu\betaarrow)\mu\delta.B_1} \\[2em]

(\mu{\mbox{-freezing)}} & \Ruled{(\mu\alphaarrow\gamma)A_1=(\mu\betaarrow\delta)B_1}{(\mu\alphaarrow)\mu\gamma.A_1=(\mu\betaarrow)\mu\delta.B_1} \\[2em]

 &\{\alphaarrow\}\subseteq\FV(A_1\to A_2)\;\; \alphaarrow\restricted A_i=\alphaiarrow=\alpha_{i_1}\ldots\alpha_{i_m}\\
{\mbox{(decomposition)}} &\{\betaarrow\}\subseteq\FV(B_1\to B_2)\;\; \betaarrow\restricted B_i=\betaiarrow=\beta_{i_1}\ldots\beta_{i_m}\\
&\Ruled{(\mu\alphaeenarrow)A_1=(\mu\betaeenarrow)B_1\;\;\;\;(\mu\alphatweearrow)A_2=(\mu\betatweearrow)B_2}
{(\mu\alphaarrow)A_1\to A_2=(\mu\betaarrow)B_1\to B_2}\\[2em]

\hline
\ear$$
\edf

Note that in the above definition $A,B$ stand for types, not $\alpha$-equivalence classes of types and $\eqalpha$ on $\tamua$ is given by
$$A\eqalpha A'\;\Rightarrow\;(\mu\alpha_1).....(\mu\alpha_n)A'\eqalpha(\mu\alpha'_1).....(\mu\alpha'_n)A'[\alpha_i :=\alpha'_i]\;\;(\alpha'_i\notin \FV(A).$$ 
Note also that in the rule ($\mu$-freezing) implicitly $\gamma\in\FV(A_1),$ $\delta\in\FV(B_1)$ and in rule (decomposition) $\{\alphaiarrow\}\subseteq\FV(A_i),$ $\{\betaiarrow\}\subseteq\FV(B_i).$\\
In a decision procedure of $\prmua a=b$ in fact these rules are used upside down. That explains the names of the rules.

\blem\label{prmuaalphacomp} Let $a\eqalpha a', b\eqalpha b'.$ Then 
$$\prmua a=b\;\Rightarrow\;\prmua a'=b'.$$
\elem
\bpf  By an easy induction on the definition of $\prmua a=b.$ \qed
\epf

So Definition \ref{prmua} defines in fact a relation on the set of $\alpha$-equivalence classes. We denote that relation also by $\prmua.$

\blem\label{freevariables}\bsub\fit $\FV(\mu\beta.A_1)=\FV(A_1[\beta :=\mu\beta.A_1])=\FV((\mu\beta)A_1).$
\item $A\redmustar B \Rightarrow \FV(A)=\FV(B).$
\item $a\redmuastar b\Rightarrow \FV(b)\subseteq\FV(a).$
\item $\prmua (\mu\alphaarrow)A = (\mu\betaarrow)B\Rightarrow |\{\alphaarrow\}| = |\{\betaarrow\}|,\; \FV((\mu\alphaarrow)A) = \FV((\mu\betaarrow)B)$ \qed
\esub
\elem

For easier use in the following we rewrite Definition \ref{prmua} for $\alpha$-equivalence classes as follows.

\bdf\label{prmuaforclasses} (Of $\prmua a=b$ for $\alpha$-equivalence classes $a,b.$)
$$\bar[b]{|ll|}
\hline\hoog{1.5em}{} 
\mbox{{(axiom)}}& {\textstyle {a=a} }\\[2em]
{\mbox{(left }}\mu{\mbox{-step)}}& \Ruled {(\mu\alphaarrow)A_1[\beta :=\mu\beta.A_1] = (\mu\alphaarrow)B}{(\mu\alphaarrow)\mu\beta.A_1 = (\mu\alphaarrow)B} 
\\[2em]
 {\mbox{(right }}\mu{\mbox{-step)}}  & \Ruled{(\mu\alphaarrow)A = (\mu\alphaarrow)B_1[\beta :=\mu\beta.B_1]}{(\mu\alphaarrow)A = (\mu\alphaarrow)\mu\beta.B_1} \\[2em]
(\mu{\mbox{-freezing)}} & \Ruled{(\mu\alphaarrow\beta)A_1=(\mu\alphaarrow\beta)B_1}{(\mu\alphaarrow)\mu\beta.A_1=(\mu\alphaarrow)\mu\beta.B_1} \\[2em]

 &\{\alphaarrow\}\subseteq\FV(A_1\to A_2)\;\; \alphaarrow\restricted A_1=\alphaeenarrow,  \alphaarrow\restricted A_2=\alphatweearrow \\
{\rm(decomposition)} &\Ruled{(\mu\alphaeenarrow)A_1=(\mu\alphaeenarrow)B_1\;\;\;\;(\mu\alphatweearrow)A_2=(\mu\alphatweearrow)B_2}
{(\mu\alphaarrow)A_1\to A_2=(\mu\alphaarrow)B_1\to B_2}\\[2em]

\hline
\ear$$
\edf

The following is an example of a derivation of $\prmua a=b.$ We reason backwards, so we write down $a=b$ and using the rules upside down we hope to finish with only axioms.

\begin{example}\label{derivationofprmua}
\bceqn
\mu\alpha\beta\gamma.((\mu\delta.\alpha)\to\gamma)&=&\mu\alpha\beta.(\alpha\to(\mu\gamma.\alpha\to\gamma))\\
&\Downarrow&\mu\rm{-freezing}\\
(\mu\alpha)\mu\beta\gamma.((\mu\delta.\alpha)\to\gamma)&=&(\mu\alpha)\mu\beta.(\alpha\to(\mu\gamma.\alpha\to\gamma))\\
&\Downarrow&\mbox{left and right } \mu\rm{-step}\\
(\mu\alpha)\mu\gamma.((\mu\delta.\alpha)\to\gamma)&=&(\mu\alpha)(\alpha\to(\mu\gamma.\alpha\to\gamma))\\
&\Downarrow&\mbox{left }\mu\rm{-step}\\
(\mu\alpha)((\mu\delta.\alpha)\to\mu\gamma.((\mu\delta.\alpha)\to\gamma))&=&(\mu\alpha)(\alpha\to(\mu\gamma.\alpha\to\gamma))\\
&\Downarrow&\rm{decomposition}\\
(\mu\alpha)(\mu\delta.\alpha)=(\mu\alpha)\alpha&&(\mu\alpha)\mu\gamma.((\mu\delta.\alpha)\to\gamma)=(\mu\alpha)(\mu\gamma.\alpha\to\gamma)\\
\mbox{left }\mu\rm{-step}~ \Downarrow~~~~~~~~~~&&~~~~~~~~~~~~~~~~~~~~~~~~~~~~\Downarrow{\rm{\mu-freezing}}\\
(\mu\alpha)\alpha=(\mu\alpha)\alpha&&~~(\mu\alpha\gamma)((\mu\delta.\alpha)\to\gamma)=(\mu\alpha\gamma)(\alpha\to\gamma)&\\
&&~~~~~~~~~~~~~~~~~~~~~~~~~~~~\Downarrow{\rm{decomposition}}\\
&&~~~(\mu\alpha)\mu\delta.\alpha=(\mu\alpha)\alpha~~~~~~~(\mu\gamma)\gamma=(\mu\gamma)\gamma\\
&&~~~~~~~~~~~~~~~~\Downarrow\mbox{left }\mu\rm{-step}\\
&&~~~~~~(\mu\alpha)\alpha=(\mu\alpha)\alpha
\eceqn
\end{example}

\brem\label{otherproofs}
Let $A=\mu\alpha\beta\gamma.((\mu\delta\alpha)\to\gamma)$ and $B=\mu\alpha\beta.(\alpha\to(\mu\gamma.\alpha\to\gamma)).$ In the example above we gave a proof of  $\prmua A=B.$ In fact one could try other proofs. For example one could start with a left $\mu$-step and a right $\mu$-step. Then one encounters at some stage again a node $A=B.$ So one could go on for ever and the result would be an infinite tree. But of course one should stop developing the sub branch at that node.\\
In general if in a derivation of $\prmua a=b$ we arrive at a node $c=d$ that we encountered already higher up in the branch from the root to that node then we stop at that node. Further on we will show that the number of nodes in the resulting tree is bounded as a function of $a,b.$\\But first we will show in the next proposition that $\prmua A=B$ iff $A$ and $B$ have a common standard reduct.
\erem

\bprop\label{prmuaiscommstred}
The following two assertions are equivalent
\bsub\item $\prmua(\mu\alphaarrow)A=(\mu\alphaarrow)B$;
\item $A$ and $B$ have a common standard reduct.
\esub
\eprop
\bpf (i)$\Rightarrow$(ii). By induction on the derivation of $\prmua(\mu\alphaarrow)A=(\mu\alphaarrow)B.$\\[1em]
(axiom)     $~~~~~~~(\mu\alphaarrow)A=(\mu\alphaarrow)A.$   This is clearly ok.\\[1em]
(left $\mu$-step)    $~~\Ruled {(\mu\alphaarrow)A_1[\beta :=\mu\beta.A_1] = (\mu\alphaarrow)B}{(\mu\alphaarrow)\mu\beta.A_1 = (\mu\alphaarrow)B} $\\[1em]
By the induction hypothsis $A_1[\beta :=\mu\beta.A_1]$ and $B$ have a common standard reduct, hence the same holds for $\mu\beta.A_1$ and $B$ by (b) of Definition \ref{standardreduction}.\\[1em]
(right $\mu$-step)      Similarly. \\[1em]
($\mu$-freezing)  $~~ \Ruled{(\mu\alphaarrow\beta)A_1=(\mu\alphaarrow\beta)B_1}{(\mu\alphaarrow)\mu\beta.A_1=(\mu\alphaarrow)\mu\beta.B_1}$ \\[1em]
By the induction hypothesis $A_1$ and $B_1$ have a common standard reduct, hence the same holds for $\mu\beta.A_1$ and $\mu\beta.B_1.$\\[1em]
$\;\;\;\;\;\;\;\;\;\;\;\;\;\;\;\;\;\;\;\;\;\;\;\;\;\;\{\alphaarrow\}\subseteq\FV(A_1\to A_2)\;\; \alphaarrow\restricted A_1=\alphaeenarrow, \;\; \alphaarrow\restricted A_2=\alphatweearrow$ \\
(decomposition)  $\Ruled{(\mu\alphaeenarrow)A_1=(\mu\alphaeenarrow)B_1\;\;\;\;(\mu\alphatweearrow)A_2=(\mu\alphatweearrow)B_2}
{(\mu\alphaarrow)A_1\to A_2=(\mu\alphaarrow)B_1\to B_2}$\\[1em]
By the induction hypothesis $A_1$ and $B_1$ have a common standard reduct and the same holds for $A_2$ and $B_2$. So it also holds for $A_1\to A_2$ and $B_1\to B_2.$\\[1em]
(ii)$\Rightarrow$ (i). By induction on the definition of $A\redst C$ and $B\redst C$ in Definition \ref{standardreduction}.\\[1em]
Case (a) for both $A$ and $B.$\\
 Then $B=A$ and we have $\prmua(\mu\alphaarrow)A=(\mu\alphaarrow)A$ by axiom.\\[1em]
Case (b) for $A$    (and similarly for $B$).\\
We have $\prmua(\mu\alphaarrow)A_1[\beta :=\mu\beta.A_1] = (\mu\alphaarrow)B$ and we get  $\prmua(\mu\alphaarrow)\mu\beta.A_1 = (\mu\alphaarrow)B$ by (left $\mu$-step).\\[1em]
Case (c) for $A$ an for $B$.\\
 Now $A=\mu\beta.A_1$ and $B=\mu\beta.B_1,\;\;  A_1\redst C_1\redstl B_1$ and therefore\\ $\mu\beta.A_1\redst \mu\beta.C_1\redstl \mu\beta.B_1.$\\
Note that $\FV(A_1) = \FV(B_1)$ by Lemma \ref{freevariables}(ii)
We distinguish two subcases\\[1em]
Subcase (c$_1$)  $\beta\in\FV(A_1)\cap\FV(B_1).$\\
By the induction hypothesis we have $\prmua(\mu\alphaarrow\beta)A_1=(\mu\alphaarrow\beta)B_1$ and hence by ($\mu$-freezing)  $\prmua(\mu\alphaarrow)\mu\beta.A_1=(\mu\alphaarrow)\mu\beta.B_1.$\\[1em]
Subcase (c$_2$)  $\beta\notin\FV(A_1),\beta\notin\FV(B_1).$\\
Now we have by the induction hypothesis $\prmua(\mu\alphaarrow)A_1=(\mu\alphaarrow).B_1$, hence we get by (left $\mu$-step) and (right $\mu$-step) 
$\prmua(\mu\alphaarrow)\mu\beta.A_1=(\mu\alphaarrow)\mu\beta.B_1.$\\[1em]
Case (d) for $A$ and for $B$.
Now we have $A_1\redst C_1\redstl B_1$,  $A_2\redst C_2\redstl B_2$ and hence  $A=A_1\to A_2\redst C_1\to C_2\redstl B_1\to B_2$. Note that by Lemma \ref {freevariables}(ii) we have $\FV(A_i)=\FV(B_i)$ and $\FV(A_1\to A_2)=\FV(B_1\to B_2).$ Let $\{\alphaarrow\}\subseteq\FV(A_1\to A_2)$ so that $(\mu\alphaarrow)A_1\to A_2,\;(\mu\alphaarrow)B_1\to B_2\in\tamua.$ Define $\alphaiarrow=\alphaarrow\restricted A_i,\; i=1,2.$ Then $(\mu\alphaiarrow)A_i,\;(\mu\alphaiarrow)B_i\in\tamua$ and by the induction hypothesis, two times, we have
$\prmua(\mu\alphaeenarrow)A_1=(\mu\alphaeenarrow)B_1,\;\prmua(\mu\alphatweearrow)A_2=(\mu\alphatweearrow)B_2.$\\[1em]
Now we have treated all cases. If we have Case (c) for $A$ and Case (a) for $B$ then in fact we have Case (c) for both $A$ and $B.$ If we have Case (d) for $A$ and Case (a) for $B$ then in fact we have Case (d) for both $A$ and $B.$
Finally Case (c) for $A$ and Case (d) for $B$  (or vice versa) cannot occur at the same time. \qed
\epf

Now we will finish the proof of decidability of $\weakeq$ by showing that the number of different nodes $c=d$ occurring in a derivation tree of $\prmua a=b$ is bounded as a function in $a,b.$\\
Note that for all these nodes $c=d$ we have $c\in\SC(a),\;d\in\SC(b)$. (See Definition \ref{redmua})\\
We start by showing that $|\SC(a)|$ is bounded as a function in $a.$

\blem\label{mualpharestricted}  $(\mu\alphaarrow)\Box((\mu\betaarrow)\Box a) =(\mu\alphaarrow\betaarrow)\Box a.$
\elem
\bpf  Immediate. Note that by the variable convention $\alphaarrow$ and $\betaarrow$ have no variables in common. \qed
\epf

\blem\label{SC} 
$\SC(\alpha)=\{\alpha\};$\\
$\SC(\mu\beta.A_1)=\{\mu\beta.A_1\}\cup(\mu\beta)\Box\SC(A_1)\cup\SC(A_1[\beta :=\mu\beta.A_1]);$\\
$\SC(A_1\to A_2)=\{A_1\to A_2\}\cup\SC(A_1)\cup\SC(A_2);$\\
$\SC((\mu\alphaarrow)A)= (\mu\alphaarrow)\Box\SC(A).$
\elem
\bpf By induction on the structure of $a\in\tamua,$ using the preceding lemma. Note that the second clause also holds for $\beta\notin\FV(A_1)$ because then $(\mu\beta)\Box\SC(A_1)=\SC(A_1) =\SC(A_1[\beta :=\mu\beta.A_1])$ by Lemma \ref{freevariables}(iii). \qed
\epf

Now $\SC([A_1[\beta :=\mu\beta.A_1])$ could be intricate but we will show in Lemma \ref{SCsimple} that we have in fact $\SC(\mu\beta.A_1)\subseteq\{\mu\beta.A_1\}\cup(\mu\beta)\Box\SC(A_1)\cup\SC(A_1)[\beta :=\mu\beta.A_1].$

\blem\label{onredmua} If $A[\beta := B]\redmuastar b$ then we have either
\bsub\item This reduction is $A[\beta := B]\redmuastar B\redmuastar b$
\item This reduction is a $[\beta := B]$ instance of $A\redmuastar a$    (So $b=a[\beta := B]$).
\esub
\elem
\bpf By induction on the length of the reduction $A[\beta := B]\redmuastar b$.\\
If the length is zero then we have (ii). Else we have $A[\beta := B]\redmuastar b'\redmua b$. The induction hypothesis for $A[\beta := B]\redmuastar b'$ gives that we have either (i) or (ii) as follows\\[1em]
(i) $A[\beta := B]\redmuastar B\redmuastar b'$  Then we have also case (i) for $A[\beta := B]\redmuastar b$.\\[1em] 
(ii) The reduction $A[\beta := B]\redmuastar b'$ is a $[\beta := B]$ instance of $A\redmuastar a'.$    So $b'=a'[\beta := B]$  We distinguish subcases for $a'$ as follows.\\[1em]
(iia)  $a'=\beta.$  Then we have Case (i) (for $A[\beta := B]\redmuastar b).$  \\[1em]
(iib)  $ a'=\alpha\neq\beta$ or $a'=(\mu\alpha).\alpha.$  These cases cannot occur because then $b'=a'$ is in nf.\\[1em]
(iic)  $a'=(\mu\alphaarrow)\mu\gamma.A'$ with $\beta\notin\{\alphaarrow,\gamma\}.$\\
In this case $(\mu\alphaarrow)$ at the root is very innocent. It is frozen and occurs in the same way at the root in all types occurring in the proof in this part (iic). Therefore for simplicity we assume $\{\alphaarrow\}=\emptyset.$ Now $b'=\mu\gamma.A'[\beta := B].$ We distinguish two subcases.\\[1em]
(iic1)  $\mu\gamma.A'[\beta := B]\redmua (\mu\gamma).(A'[\beta := B])$          ($\ast$)\\
Then $\gamma\in\FV(A'[\beta := B])$, hence also $\gamma\in\FV(A'),$ because $\gamma\notin\FV(B)$ by the variable convention. So ($\ast$) is a  $[\beta := B]$ instance of $\mu\gamma.A'\redmua (\mu\gamma).A'$. In total we have that $A[\beta := B]\redmuastar B'\redmua b$ is an instance of $A\redmuastar\mu\gamma.A'\redmua (\mu\gamma).A'$ and we are in Case (ii). \\[1em]
(iic2) $b'=\mu\gamma.A'[\beta := B]\redmua A'[\beta := B][\gamma :=\mu\gamma.A'[\beta := B]] = b.$\\
By the substitution lemma we get $b'=\mu\gamma.A'[\beta := B]\redmua A'[\gamma :=\mu\gamma.A'][\beta := B] =b.$ \\
Now we have that $A[\beta := B]\redmuastar a'[\beta := B]\redmua a[\beta := B]$ is an instance of $A\redmuastar a'=\mu\gamma.A'\redmua a=A'[\gamma :=\mu\gamma.A']$ and we are in Case (ii).\\[1em]
(iid) $a'=(\mu\alphaarrow)(A_1\to A_2).$ Then $b'=(\mu\alphaarrow)(A_1[\beta := B]\to A_2[\beta := B])\redmua (\mu\alphaarrow)\Box(A_i[\beta := B]) = b$ and $a'\redmua (\mu\alphaarrow)\Box A_i$. Now we are in case (ii) again.
\epf

\blem\label{SCsimple}   $\SC(\mu\beta.A_1)\subseteq\{\mu\beta.A_1\}\cup(\mu\beta)\Box\SC(A_1)\cup\SC(A_1)[\beta :=\mu\beta.A_1].$
\elem
\bpf Let $\mu\beta.A_1\redmuastar b$ be a reduction path of minimal length.\\
If the length is zero then $b=\mu\beta.A_1.$ Now let the length be bigger then zero. We distinguish two cases.\\[1em]
Case (i) $\mu\beta.A_1\redmua (\mu\beta).A_1\redmuastar b$\\
Now $\beta\in\FV(A_1)$ and $b\in\SC((\mu\beta).A_1)=(\mu\beta)\Box\SC(A_1).$\\[1em]
Case(ii)  $\mu\beta.A_1\redmua A_1[\beta :=\mu\beta.A_1]\redmuastar b.$\\
We distinguish two subcases according to Lemma \ref{onredmua} \\
(iia) $\mu\beta.A_1\redmua A_1[\beta :=\mu\beta.A_1]\redmuastar\mu\beta.A_1\redmuastar b.$  This cannot occur because it is not a reduction of minimal length.\\
(iib) $A_1[\beta := \mu\beta.A_1]\redmuastar b$ is a $[\beta := \mu\beta.A_1]$ instance of $A_1\redmuastar a.$ Now $a\in\SC(A_1)$ hence $b=a[\beta :=\mu\beta.A_1]\in\SC(A_1)[\beta :=\mu\beta.A_1].$ \qed
\epf

\bdf\label{length} \bceqn
l(\alpha)&=&1;\\
l(A_1\to A_2)&=&1 + l(A_1) + l(A_2);\\
l(\mu\beta.A_1)&=&1 + l(A_1);\\
l((\mu\alphaarrow).A)&=&l(A).\eceqn
\edf

\blem\label{lengthSC} $|\SC((\mu\alphaarrow)A)|=|\SC(A)|\leq 3^{l(A)}.$
\elem
\bpf By Lemmas \ref{SC} and \ref{SCsimple}. \qed
\epf

\blem\label{numberofnodes} Let $n=n((\mu\alphaarrow)A=(\mu\alphaarrow)B)$ be the number of different nodes in a complete derivation treee for $\prmua(\mu\alphaarrow)A=(\mu\alphaarrow)B.$ Then 
$$n\leq 3^{(3^{l(A)+l(B)})}.$$
\elem
\bpf Easy. Note that each node in the derivation tree has at most three daughters. \qed
\epf

\bcor\bsub\fit $\prmua(\mu\alphaarrow)A=(\mu\alphaarrow)B$ is decidable.
\item $\prmua A=B$ is decidable. \qed
\esub
\ecor

\bcor $A\weakeq B$ is decidable.
\ecor

\bpf  By Remark \ref{facts} and Proposition \ref{prmuaiscommstred}. \qed
\epf

\section {A decision procedure of weak $\mu$-equality  for types themselves}

In this subsection we work with the $\mu$-types themselves instead of their $\alpha$-equivalence classes. On this set of types $\redmualpha$ is the reduction relation on the set of types $\tamu$ generated by $\redmu$ and $\redalpha$. We show that the relation $\weakmualphaeq$ on $\tamu$ is decidable via the first order decision procedure of Definition \ref{prmua} for types.

\bdf $\redmualpha$ is the reduction relation on the set of types $\tamu$ generated by $\redmu$ and $\redalpha.$\\ $\weakmualphaeq$ is the transitive, symmetric, reflexive closure of $\redmualpha.$
 $\redmualphastar$ denotes zero or more $\redmualpha$ steps.
\edf

We will show 
$$A\weakmualphaeq B \Leftrightarrow \;\prmua A=B.$$
where $\prmua$ derives equations between types in $\tamua$, as given in Definition \ref{prmua} and moreover that this relation is decidable.\\
From now on  $\alpha$-quivalence classes in $\tamu$ and $\tamua$ are denoted by $[A]$ and $[a].$ $\weakeqcl$ stands for equality on the set  of $\alpha$-equivalence classes $[A]$ and 
$\prmuacl$ denotes the derivation system of Definition \ref{prmuaforclasses} for classes $[a]$ where $a\in\tamua.$

\blem\label{prmuaisprmuacl} $\prmua a=b \Leftrightarrow \;\prmuacl [a]=[b]$ and both derivations can be done by the "same" steps. Equivalently they have the same length.
\elem
\bpf Immediate \qed
\epf

\blem\label{eqmualphaiseqmu} $A\weakmualphaeq B \Leftrightarrow [A]\weakeqcl [B].$
\elem
\bpf Immediate \qed
\epf

Combining Remark \ref{facts}, Proposition \ref{prmuaiscommstred} and Lemmas \ref{prmuaisprmuacl} and \ref{eqmualphaiseqmu} we get

\bcor  $A\weakmualphaeq B \Leftrightarrow \;\prmua A=B.$ \qed
\ecor

Finally by Lemma \ref{prmuaisprmuacl} we have that $\prmua A=B$ is decidable. In fact there is a bijection from the derivation tree for $\prmua A=B$ on the derivation tree for $\prmuacl [A]= [B].$ So we have the following

\bth $A\weakmualphaeq B$ is decidable via the first order decision procedure $\prmua A=B.$ \qed
\eth

\brem In this part 2 we do not have Lemma \ref{SCsimple} for types. It only holds modulo $\alpha$-equivalence.
\erem

\newenvironment{References} {\begin{center}\textbf{References}\end{center} \begin{quote}} {\end{quote}}
\begin{References}
[1] Felice Cardone and Mario Coppo. Decidability Properties of Recursive Types. In C. Blundo and C. Laneve, editors, {\it ICTCS 2003} volume 2841 of {\it LNCS,} pages 242-255. Springer, 2003.\\[1em]
[2]J\"org Endrullis, Clemens Grabmayer, Jan Willem Klop, Vincent van Oostrom. On equal $\mu$-terms. 2011. Free university Amsterdam.\\[1em]
[3] V. van Oostrom. FD \`a la Melli\`es. February 1997. Free  University Amsterdam.
\end{References}
\end{document}